\pdfoutput=1
\documentclass[]{llncs}

\usepackage[utf8]{inputenc}
\usepackage{amsmath,amssymb,amsfonts,mathrsfs,stmaryrd,wasysym}

\usepackage[usenames,svgnames,table]{xcolor}
\usepackage{multirow}

\usepackage{url}

\usepackage{paralist}

\usepackage{setspace}
\usepackage{calc}

\usepackage{graphicx}

\usepackage{hyperref}
\usepackage{xspace}

\usepackage[normalem]{ulem}

\usepackage{booktabs}

\usepackage[text size=small, backgroundcolor=blue!30,bordercolor=blue,linecolor=blue]{todonotes}
\newcommand{\myparagraph}[1]{\vspace{1.5ex}\noindent\textbf{#1}\hspace{1.5ex}}

\newcommand{\swat}{SWaT\xspace}
\newcommand{\cps}{CPS\xspace}

\begin{document}

\title{Towards Learning and Verifying Invariants of Cyber-Physical Systems by Code Mutation}

\author{Yuqi Chen \and Christopher M. Poskitt \and Jun Sun}

\institute{Singapore University of Technology and Design, Singapore}

\maketitle

\begin{abstract}
  Cyber-physical systems (\cps), which integrate algorithmic control with physical processes, often consist of physically distributed components communicating over a network. A malfunctioning or compromised component in such a \cps can lead to costly consequences, especially in the context of public infrastructure. In this short paper, we argue for the importance of constructing invariants (or models) of the physical behaviour exhibited by \cps, motivated by their applications to the control, monitoring, and attestation of components. To achieve this despite the inherent complexity of \cps, we propose a new technique for learning invariants that combines machine learning with ideas from mutation testing. We present a preliminary study on a water treatment system that suggests the efficacy of this approach, propose strategies for establishing confidence in the correctness of invariants, then summarise some research questions and the steps we are taking to investigate them.
\end{abstract}

\section{Introduction}

Cyber-physical systems~(\cps), characterised by their tight integration of algorithmic control and physical processes, are prevalent across engineering domains as diverse as aerospace, autonomous vehicles, and medical monitoring; they are also used to control critical public infrastructure such as smart grids and water treatment plants~\cite{Khaitan-McCalley15a,Lee08a}. In such contexts, \cps often consist of distributed software components (the ``cyber'' part) that communicate over a network and interact with their local environments via sensors and actuators (the ``physical'' part). A component that exhibits faulty behaviour---or worse still, becomes compromised~\cite{Cardenas-Amin-Sastry08a}---can lead to costly and damaging consequences, motivating research into approaches for ensuring their correctness, safety, and security.

Reasoning about a \cps as a whole, however, is very challenging, given that models must capture both discrete behaviour in the cyber part as well as continuous behaviour in the physical part~\cite{Zheng-et_al15a}. With source code for the former and ordinary differential equations~(ODEs) for the latter, it becomes possible to model the \cps as a hybrid system and apply a variety of techniques (e.g.~model checking~\cite{Frehse-et_al11a}, SMT solving~\cite{Gao-Kong-Clarke13a}, non-standard analysis~\cite{Hasuo-Suenaga12a}, concolic testing~\cite{Kong-et_al16a}, or theorem proving~\cite{Platzer-Quesel08a,Quesel-et_al16a}). Yet \cps are inherently complex, and even with domain-specific expertise, it can be difficult to determine ODEs that are accurate enough in practice: there might always remain some discrepancy between the verified model and the behaviour of the actual \cps, emphasising the importance of techniques that can be applied at runtime~\cite{Mitsch-Platzer14a}.

\myparagraph{Our Approach.} As an alternative to the endeavour of manual modelling, we pursue in this paper a more systematic approach. We propose to apply machine learning~(ML) to the sensor data of \cps to construct models in the form of \emph{invariants}---conditions that must hold in all states amongst the physical processes controlled by the \cps---and to make those invariants checkable at runtime. To achieve this, the learner must be trained on traces of sensor data representing ``normal'' runs (the positive case, satisfying the invariant), and also on traces representing incorrect behaviour (the negative case); the former being easy to obtain, but the latter requiring more ingenuity. We obtain our negative traces by the novel application of code mutation (\emph{\`{a} la} mutation testing~\cite{Jia-Harman11a}) to the software components of \cps. Besides characterising the \cps, the learnt invariants have some important applications in controlling, monitoring, and attesting the software components~\cite{Roth-McMillin13a}. It is thus important to ascertain that the learnt invariants \emph{actually} are invariants of the \cps: to address this, we propose to verify them using statistical model checking and symbolic execution.

\myparagraph{Our Contributions.} This short paper describes a novel approach for generating invariants (or models) of CPS, based on the application of machine learning to traces of sensor data obtained under mutated software components. We present the results of a preliminary experiment on (a simulator of) Secure Water Treatment~(\swat)~\cite{SWaT-Reference}, a water purification testbed, which suggest the efficacy of the approach and motivate the need for further research. Furthermore, we propose the use of statistical model checking and symbolic execution for establishing confidence in the correctness of the learnt invariants, and highlight some important open research questions which we are investigating in ongoing work. For the formal methods community, this paper represents the start of a line of work to model and verify---``warts and all''---a complex, real-world CPS. For the CPS community, it describes a systematic approach for constructing invariants that can be applied in controlling, monitoring, and attesting software components. For the ML community, it presents a new application of learning arising from a novel combination of ideas from CPS and mutation testing.

\section{\swat Testbed and Cyber-Physical System Invariants}

\myparagraph{\swat.} We are currently investigating our approach in the context of a particular \cps: the \swat testbed~\cite{SWaT-Reference}. \swat, built for cyber-security research at the Singapore University of Technology and Design, is a scaled-down but fully operational water treatment plant, capable of producing five gallons of safe drinking water per minute. Water is treated in six distinct, co-operating stages, under which it undergoes chemical processes such as ultrafiltration, de-chlorination, and reverse osmosis. Each stage is controlled by an independent programmable logic controller~(PLC), which receives sensor data such as water flow rates and tank levels, and then computes signals to send to actuators including pumps and motorised valves. This communication all takes place over a network. Sensor data is also available to a Supervisory Control and Data Acquisition~(SCADA) system, and is recorded by a historian to facilitate offline analyses.

Control is expressed in the programs that PLCs repeatedly cycle through. These are structurally very simple, essentially boiling down to big (nested) if-statements. The programs use only the simplest constructs: loops, for example, are completely absent. Furthermore, the source code can easily be viewed, modified, and re-deployed to the PLCs using Rockwell's RSLogix 5000, an industrial-standard software suite. While the cyber part of \swat is thus relatively simple, the same is not true of the physical part: runs of the system are governed by laws concerning the dynamics of water flow, the evolution of pH values, and the chemical processes associated with the six water treatment stages.

To complement the \swat testbed, we also have access to a simulator implemented in Python (relying on some of its scientific libraries). The cyber part is simulated faithfully as the PLC code was translated to Python directly. Since the actual ODEs governing the physical part of \swat are unknown, the simulator is not as accurate in this regard. The ODEs it does implement, however, have been improved over time by cross-validating data from the simulator with real \swat data collected by the historian.

\vspace{-2pt}\myparagraph{CPS Invariants.} The safety of water treatment plants is of paramount importance, as breaches or malfunctioning components can lead to costly consequences. In \swat, for example, there is a risk of damaging the mechanics of the system if the water levels in certain tanks become too high or too low~\cite{Kang-et_al16a}. One way to detect when runs of a system are diverging into such territory is to monitor invariants---conditions that must hold in all states amongst the physical processes controlled by the CPS---and raise an alarm when they are no longer satisfied. This approach has been applied to a number of \cps~\cite{Choudhari-et_al13a,Paul-et_al14a}, including for stages of \swat itself~\cite{Adepu-Mathur16b,Adepu-Mathur16a}. Typically, however, the invariants are \emph{manually} derived using the laws of physics and domain-specific knowledge. Moreover, they are derived for specific, expected physical relationships, and may not capture other important patterns hiding in the sensor data.

Beyond providing a characterisation of \cps and their important applications in monitoring for safety, invariants can also be seen as facilitating a form of code attestation. That is to say, if the actual behaviour of a \cps does not satisfy our mathematical model of the physical world under its control (i.e.~the invariant), then it is possible that the cyber part has been compromised and that ill-intended manipulations are occurring. This form of attestation is known as \emph{physical attestation}~\cite{Roth-McMillin13a,Valente-Barreto-Cardenas14a}, and while weaker than typical software- and hardware-based attestation schemes (e.g.~\cite{Alves-Felton04a,Anati-et_al13a,Castelluccia-et_al09a,Seshadri-et_al04a}), it is much more lightweight---neither the firmware nor the hardware of the PLCs require modification.

\section{Learning with Mutants}
\label{sec:learning_with_mutants}

\myparagraph{Learning \swat Invariants.} Rather than deriving further invariants for \swat manually, we propose to learn them systematically by applying ML---initially, Support Vector Machines~(SVM)---to traces of \swat sensor data, taking the \emph{classifiers} they learn as our invariants. To learn such a classifier, SVM must be provided with traces that should be classified as positive (i.e.~correct behaviour) and traces that should be classified as negative. The data available from the \swat historian can be seen as representing correct (and thus positive) behaviour of the system as a whole: the \swat PLCs and actual (unknown) ODEs together. In contrast, we propose to collect negative traces by running the system under small manipulations. Since we cannot change the ODEs (we cannot yet bend the laws of physics!), we propose to manipulate the part of \swat that we can: the programs running on the PLCs.

As previously discussed, it is straightforward to change the PLC programs of \swat and collect some negative traces, but it is more challenging to do so in a systematic way that ensures the strength of the invariant and precision of the classifier. The solution we propose is directly inspired by mutation testing~\cite{Jia-Harman11a}, a fault-based testing technique that deliberately seeds errors---small, syntactic changes called \emph{mutations}---into multiple copies of a program, which are executed to assess the quality of a test suite (good ones should detect the mutants). Rather than using mutations to improve the completeness of a test suite, we are using them to generate a more comprehensive set of negative traces for training on. By training on traces resulting from small syntactic changes, we hope to learn a classifier as close to the boundary between correct and suspicious behaviour as possible. Our rationale is that smaller changes are more likely to reveal negative traces that are relevant in practice, corresponding, for example, to isolated PLCs or sensors failing, or an attacker attempting to keep their changes undetected.

Using mutations for learning is also attractive because of the structural simplicity of the PLC programs. Were we assessing a test suite on them, we could do so efficiently and without redundancy by using the five basic (arithmetic, relational, and logical) mutation operators identified by Offutt et al.~\cite{Offutt-et_al96a}. We hypothesise that (and are investigating whether) this result has an analogue for learning that could help us in minimising the number of redundant traces. Even if so, there remain some additional challenges to overcome. For example, if mutations are not executed, this must be detected, and thus the traces rejected as negative samples. Even if a mutant is executed, it may not lead to a physical effect immediately (or ever) and thus could generate traces indistinguishable from positive ones. Other issues include how many mutations to use in each copy, and how to handle valid modes of operation in \swat that are rarely entered.

\myparagraph{Preliminary Evaluation.} As a very first step towards evaluating the outlined approach, we undertook an experiment to ascertain the effectiveness of a classifier learnt from traces produced by the \swat simulator under a number of manually applied mutations. Note that we used the simulator to facilitate a quick proof-of-concept without the resource costs of the real system (e.g.~water usage, human monitoring); this ML approach can be applied to traces collected from the real system in the same way.

First, we manually launched the \swat simulator in three different initial states (i.e.~assignments of variables modelling sensors), collecting three traces of correct behaviour each spanning 30 minutes. Following this, we made 20 copies of the PLC code and manually applied a different (random) mutation to each. Of these 20 mutants, 14 of them generated traces equivalent to correct behaviour and were manually rejected. Seven mutants generated different traces, although one mutant was rejected for generating a trace too similar to another. The six remaining mutants were selected to generate our negative traces; three of the mutations each modified an assignment, whereas the other three modified an arithmetic expression in a conditional guard. We generated traces for each mutant using the same three initial states as before.

We proceeded to apply SVM to learn six classifiers for the six mutants respectively, each against the correct code. We selected 10 features: the first five representing the water levels of the five tanks, and the next five representing the same levels after 250ms. For training the classifiers and evaluating their accuracy, we applied $k$-fold cross-validation to the traces with $k=5$. On average, the classifiers achieved an accuracy of 99\%.

Finally, we applied SVM to all the traces from all six mutants to learn a single classifier, i.e.~to determine whether a trace represents correct behaviour or the behaviour caused by any one of the mutations. We found that this combined classifier maintained a similar level of accuracy to the individual ones: $98.41\%$. We extracted the learnt invariant from this classifier, which, albeit complicated, expresses a linear relationship between water tank levels (mm) at one time point ($v_1,\dots v_5$) and 250ms after it ($v'_1,\dots v'_5$). For simplicity of presentation, the coefficients are given below to three decimal places. The full model and training data are all available online (see~\cite{Supplementary-Material}).

\noindent \begin{center}\begin{tabular}{l}
\small $-0.349v_1 + 9.789v_2 -10.192v_3 + 0.803v_4 -5.561v_5$\\
\small $-0.630v'_1 -10.455v'_2 + 10.333v'_3 + 0.803v'_4 + 3.928v'_5\ < -786.416$
\end{tabular}\end{center}

This experiment has shown that it is possible to apply SVM to learn an accurate classifier for traces of sensor data, using the negative samples generated under a small number of mutated PLC programs. It is, of course, too limited in its present scope to allow for more general conclusions; a much more extensive evaluation of the outlined approach is needed, and is underway. It does however suggest the feasibility of the basic idea, and has highlighted a number of important challenges. For instance, the process should be more automatic: mutation operators should be applied automatically, as should the detection of unexercised mutations, as well as the comparison of the generated traces against the positive ones. Furthermore, to ensure as strong an invariant and precise a classifier as possible, a number of questions must be answered empirically, regarding, e.g.~the number of mutations (and the possibility of multiple mutations per copy), the sufficiency of mutation operators, and the length of traces. 

Our experiment also highlighted the role that a simulator can play in mutation ``screening'' before applying them to the real \swat system and collecting negative traces that are based on the actual ODEs. This helps to avoid wasting time and resources otherwise lost by applying the mutations to the real PLCs first. Note that while the ML technique can be applied to \swat data in exactly the same way as for the simulator, a human technician must be present while collecting the data itself to ensure that the mutations do not lead the system into a state that causes damage. This raises another research question: whether one can determine a class of ``safe'' mutations for \swat that still facilitate a precise classifier but avoid entirely the possibility of causing damage.

\section{Correctness of Invariants}
\label{sec:correctness_invariant}

Our preliminary experiment has allowed us to learn a new invariant for \swat (or rather, at least to begin with, its simulator). But is it \emph{actually} an invariant? It is not particularly intuitive to reason about. And even if it were, to argue for its correctness, we would need some expertise in the physics of water treatment plants; a requirement we wanted to avoid in the first place. As alternatives to manual, ad hoc proofs, we propose two contrasting approaches for establishing confidence in the correctness of invariants, and highlight their well-suitedness to CPS like \swat.

First, we will apply statistical model checking~(SMC) to \swat, a standard technique for analysing and verifying CPS~\cite{Clarke-Zuliani11a}. In SMC, executions of the system (i.e.~traces of sensor data) are observed, and hypothesis testing or statistical estimation techniques are applied to determine whether or not the executions provide statistical evidence of the invariant holding. SMC estimates the probability of correctness, rather than guaranteeing it outright, but is simple to apply to \swat (and its simulator) since it only requires the system to be executable. Furthermore, should the ODEs of the \swat simulator become more accurate in the future, then our mutation-based learning approach could take place entirely on that; SMC could then determine whether or not the learnt invariants are also invariants of the real system, without having to apply any mutations to it.

Second, we will investigate the use of symbolic execution for analysing \swat with respect to a learnt invariant. In the PLC programs, symbolic values will be used to abstract away from concrete sensor inputs. The technique will then build, along the different paths of the PLC code, path constraints over the symbolic values (i.e.~path conditions in conjunction with an assertion based on the learnt invariant). The PLC programs have a simple structure that is well-suited to this task: they are free of loops, and the paths through the programs are short (maximal depth of three; maximal branching of 28). Our invariants, however, are based upon sensor readings at two different time points, so we cannot analyse them with respect to the cyber part of \swat alone: a model of the physical processes is needed too, for reasoning about the effects that signals will have. As we have discussed, we cannot expect to manually derive a completely accurate one, but we could nonetheless use approximate models (e.g.~as defined in the simulator), or even models of \swat that were automatically constructed using different approaches to ours (e.g.~the probabilistic model of~\cite{Wang-et_al16a}).

It should be emphasised that while neither technique can fully guarantee correctness, they differ in where precision is lost, and so should complement each other in helping to establish confidence in the learnt invariants. SMC, for example, estimates a probability of correctness based only on the executions it is provided with (leading to challenges such as handling rare events); yet by working with actual system executions, its results are based on the actual physical processes. Symbolic execution, in contrast, must work with an approximate physical model, but performs an analysis on the actual source code in the cyber part (and not just on a subset of the possible system executions).

\section{Conclusion and Next Steps}

This short paper has proposed a novel approach for learning invariants of CPS that trains a ML technique such as SVM on positive and negative traces of sensor data, with the latter obtained by applying mutation operators to copies of the programs in the cyber part---the part of the CPS that we can most easily control. We presented a preliminary study on \swat, a raw water treatment plant, that suggested the effectiveness of constructing invariants this way. We furthermore outlined the use of SMC and symbolic execution for establishing confidence in the correctness of learnt invariants, and discussed their use in CPS applications such as physical attestation.

Much work remains to be done to truly ascertain the effectiveness of our approach for \cps. First, we will automate---as much as possible---our experiment on the \swat simulator, to allow for classifiers to be trained on several additional mutants and initial states more easily, and to automatically detect those mutants that do not cause the system to exhibit different physical behaviour. Then, within this framework, we will begin investigating the challenges raised in Section~\ref{sec:learning_with_mutants} and the verification approaches outlined in Section~\ref{sec:correctness_invariant}, before shifting our experimentation to traces obtained from the real \swat system. We will investigate the use of ML systems other than SVM, and compare our supervised model learning approach against proposed unsupervised ones for \cps (e.g.~\cite{Maier14a,Vodencarevic-et_al11a}). Finally, we will investigate the application of learnt invariants to code attestation, by instigating cyber-attacks on the \swat system and evaluating whether or not our classifiers are effective in detecting them. \\

\noindent\emph{Acknowledgements.} We thank Pingfan Kong for assisting us with the \swat simulator, and the anonymous referees for their helpful comments and criticisms. This work was supported by NRF Award No.\ NRF2014NCR-NCR001-40.

\bibliographystyle{splncs03}

\bibliography{references}

\begin{thebibliography}{10}
\providecommand{\url}[1]{\texttt{#1}}
\providecommand{\urlprefix}{URL }

\bibitem{SWaT-Reference}
{Secure Water Treatment (SWaT)}.
  \url{http://itrust.sutd.edu.sg/research/testbeds/secure-water-treatment-swat/},
  acc.: September\ 2016

\bibitem{Supplementary-Material}
Supplementary material. \url{http://sav.sutd.edu.sg/?page_id=3258}

\bibitem{Adepu-Mathur16b}
Adepu, S., Mathur, A.: Distributed detection of single-stage multipoint cyber
  attacks in a water treatment plant. In: Proc.\ {ACM} Asia Conference on
  Computer and Communications Security (AsiaCCS 2016). pp. 449--460. {ACM}
  (2016)

\bibitem{Adepu-Mathur16a}
Adepu, S., Mathur, A.: Using process invariants to detect cyber attacks on a
  water treatment system. In: Proc.\ International Conference on ICT Systems
  Security and Privacy Protection (SEC 2016). IFIP AICT, vol. 471, pp. 91--104.
  Springer (2016)

\bibitem{Alves-Felton04a}
Alves, T., Felton, D.: {TrustZone}: Integrated hardware and software security.
  {ARM} white paper (2004)

\bibitem{Anati-et_al13a}
Anati, I., Gueron, S., Johnson, S.P., Scarlata, V.R.: Innovative technology for
  {CPU} based attestation and sealing. Intel white paper (2013)

\bibitem{Cardenas-Amin-Sastry08a}
C{\'{a}}rdenas, A.A., Amin, S., Sastry, S.: Research challenges for the
  security of control systems. In: Proc.\ USENIX Workshop on Hot Topics in
  Security (HotSec 2008). {USENIX} Association (2008)

\bibitem{Castelluccia-et_al09a}
Castelluccia, C., Francillon, A., Perito, D., Soriente, C.: On the difficulty
  of software-based attestation of embedded devices. In: Proc.\ ACM Conference
  on Computer and Communications Security (CCS 2009). pp. 400--409. ACM (2009)

\bibitem{Choudhari-et_al13a}
Choudhari, A., Ramaprasad, H., Paul, T., Kimball, J.W., Zawodniok, M.J.,
  McMillin, B.M., Chellappan, S.: Stability of a cyber-physical smart grid
  system using cooperating invariants. In: Proc.\ IEEE Computer Software and
  Applications Conference (COMPSAC 2013). pp. 760--769. IEEE (2013)

\bibitem{Clarke-Zuliani11a}
Clarke, E.M., Zuliani, P.: Statistical model checking for cyber-physical
  systems. In: Proc.\ International Symposium on Automated Technology for
  Verification and Analysis (ATVA 2011). LNCS, vol. 6996, pp. 1--12. Springer
  (2011)

\bibitem{Frehse-et_al11a}
Frehse, G., Guernic, C.L., Donz{\'{e}}, A., Cotton, S., Ray, R., Lebeltel, O.,
  Ripado, R., Girard, A., Dang, T., Maler, O.: {SpaceEx}: Scalable verification
  of hybrid systems. In: Proc.\ International Conference on Computer Aided
  Verification (CAV 2011). LNCS, vol. 6806, pp. 379--395. Springer (2011)

\bibitem{Gao-Kong-Clarke13a}
Gao, S., Kong, S., Clarke, E.M.: {dReal}: An {SMT} solver for nonlinear
  theories over the reals. In: Proc.\ International Conference on Automated
  Deduction (CADE 2013). LNCS, vol. 7898, pp. 208--214. Springer (2013)

\bibitem{Hasuo-Suenaga12a}
Hasuo, I., Suenaga, K.: Exercises in nonstandard static analysis of hybrid
  systems. In: Proc.\ International Conference on Computer Aided Verification
  (CAV 2012). LNCS, vol. 7358, pp. 462--478. Springer (2012)

\bibitem{Jia-Harman11a}
Jia, Y., Harman, M.: An analysis and survey of the development of mutation
  testing. IEEE Transactions on Software Engineering  37(5),  649--678 (2011)

\bibitem{Kang-et_al16a}
Kang, E., Adepu, S., Jackson, D., Mathur, A.P.: Model-based security analysis
  of a water treatment system. In: Proc.\ International Workshop on Software
  Engineering for Smart Cyber-Physical Systems (SEsCPS 2016). pp. 22--28. ACM
  (2016)

\bibitem{Khaitan-McCalley15a}
Khaitan, S.K., McCalley, J.D.: Design techniques and applications of
  cyberphysical systems: A survey. {IEEE} Systems Journal  9(2),  350--365
  (2015)

\bibitem{Kong-et_al16a}
Kong, P., Li, Y., Chen, X., Sun, J., Sun, M., Wang, J.: Towards concolic
  testing for hybrid systems. In: Proc.\ International Symposium on Formal
  Methods (FM 2016). LNCS-FM, Springer (2016), to appear

\bibitem{Lee08a}
Lee, E.A.: Cyber physical systems: Design challenges. In: Proc.\ International
  Symposium on Object-Oriented Real-Time Distributed Computing (ISORC 2008).
  pp. 363--369. IEEE (2008)

\bibitem{Maier14a}
Maier, A.: Online passive learning of timed automata for cyber-physical
  production systems. In: Proc.\ {IEEE} International Conference on Industrial
  Informatics (INDIN 2014). pp. 60--66. {IEEE} (2014)

\bibitem{Mitsch-Platzer14a}
Mitsch, S., Platzer, A.: {ModelPlex}: Verified runtime validation of verified
  cyber-physical system models. In: Proc.\ International Conference on Runtime
  Verification (RV 2014). LNCS, vol. 8734, pp. 199--214. Springer (2014)

\bibitem{Offutt-et_al96a}
Offutt, A.J., Lee, A., Rothermel, G., Untch, R.H., Zapf, C.: An experimental
  determination of sufficient mutant operators. ACM Transactions on Software
  Engineering and Methodology (TOSEM)  5(2),  99--118 (1996)

\bibitem{Paul-et_al14a}
Paul, T., Kimball, J.W., Zawodniok, M.J., Roth, T.P., McMillin, B.M.,
  Chellappan, S.: Unified invariants for cyber-physical switched system
  stability. IEEE Transactions on Smart Grid  5(1),  112--120 (2014)

\bibitem{Platzer-Quesel08a}
Platzer, A., Quesel, J.: {KeYmaera}: A hybrid theorem prover for hybrid systems
  (system description). In: Proc.\ International Joint Conference on Automated
  Reasoning (IJCAR 2008). LNCS, vol. 5195, pp. 171--178. Springer (2008)

\bibitem{Quesel-et_al16a}
Quesel, J., Mitsch, S., Loos, S.M., Arechiga, N., Platzer, A.: How to model and
  prove hybrid systems with {KeYmaera}: a tutorial on safety. International
  Journal on Software Tools for Technology Transfer  18(1),  67--91 (2016)

\bibitem{Roth-McMillin13a}
Roth, T.P., McMillin, B.M.: Physical attestation of cyber processes in the
  smart grid. In: Proc.\ International Workshop on Critical Information
  Infrastructures Security (CRITIS 2013). LNCS, vol. 8328, pp. 96--107.
  Springer (2013)

\bibitem{Seshadri-et_al04a}
Seshadri, A., Perrig, A., van Doorn, L., Khosla, P.K.: {SWATT:} {SoftWare-based
  ATTestation} for embedded devices. In: Proc.\ IEEE Symposium on Security and
  Privacy (S{\&}P 2004). p. 272. IEEE (2004)

\bibitem{Valente-Barreto-Cardenas14a}
Valente, J., Barreto, C., C{\'{a}}rdenas, A.A.: Cyber-physical systems
  attestation. In: Proc.\ IEEE International Conference on Distributed
  Computing in Sensor Systems (DCOSS 2014). pp. 354--357. IEEE (2014)

\bibitem{Vodencarevic-et_al11a}
Vodencarevic, A., Kleine{ }B{\"{u}}ning, H., Niggemann, O., Maier, A.:
  Identifying behavior models for process plants. In: Proc.\ IEEE Conference on
  Emerging Technologies {\&} Factory Automation (ETFA 2011). pp. 1--8. IEEE
  (2011)

\bibitem{Wang-et_al16a}
Wang, J., Sun, J., Yuan, Q., Pang, J.: Should we learn probabilistic models for
  model checking? {A} new approach and an empirical study. CoRR  abs/1605.08278
  (2016), \url{http://arxiv.org/abs/1605.08278}

\bibitem{Zheng-et_al15a}
Zheng, X., Julien, C., Kim, M., Khurshid, S.: Perceptions on the state of the
  art in verification and validation in cyber-physical systems. IEEE Systems
  Journal  PP(99),  1--14 (2015)

\end{thebibliography}

\end{document}